\documentstyle[epsfig,aps]{revtex}
\begin{document}

\title{\bf Search for delays between unlike particle emissions in
relativistic heavy-ion collisions }

\vspace{0.2in}

\author{R. Lednicky$^1$, S. Panitkin$^2$,  and Nu Xu$^2$}

\address{1 Institute of Physics, Na Slovance 2, 18221 Prague 8, Czech Republic}
\address{2 Nuclear Science Division, LBNL, Berkeley, CA 94720, USA}

\vspace{0.2in}

\maketitle

\begin{abstract}
The possibilities of unlike particle
correlations for a study of particle delays and spatial shifts
in relativistic heavy ion collisions
are demonstrated.
This report represents the unpublished Ref. [4] in the paper by
S. Voloshin et al., Phys. Rev. Lett. {\bf 79} (1977) 4766;
the report abstract was submitted to the conference Quark Matter'97.
\end{abstract}

\vspace{0.25in}

One of the major purposes in studying
heavy-ion collisions is to understand
the physics of particle interactions at both high
particle and high energy densities and its dependence on the colliding
energy.
Thus, based on the
free space cross section (or mean-free-path) arguments,
it was suggested long time ago \cite{nag82}
that the freeze-out sequence is: photons, K$^+$'s, protons and
pions.
The transport model calculations then confirmed this expectation
except for protons which appear to be emitted latest in the mean.
This is demonstrated in Fig. 1 where we present the particle
yields as functions of the emission time in the reaction c.m.s.
for $Pb + Pb$ central collisions at 158 A$\cdot$GeV simulated with the
RQMD (v2.3) code.
Nowadays, in connection with the experiments at SPS with heaviest
nuclei and the future experiments at RHIC and LHC, there are
of special interest the effects of the eventual phase
transitions, like the production of quark-gluon plasma or strangeness
distillation, leading to specific delays in particle production.
Therefore the measurement of the freeze-out sequence may be able to
provide crucial information on the properties of the fireball produced
in heavy-ion collisions.

In this report, we address the problem of relative
delays and spatial shifts among unlike particles.
We use the transport RQMD (v2.3) code to generate the
158 A$\cdot$GeV Pb+Pb central collisions. The correlation functions for
unlike particle pairs: ($\pi^+, K^+$), ($\pi^+,p$) and ($K^+,p$) are
then calculated with the final state interaction code \cite{led82,gmi86}.

In order to minimize the effect of a fast longitudinal motion of the
particle sources, we consider here only the mid--rapidity region
$|y|<1$ and calculate the two--particle correlation functions in the
longitudinally co-moving system (LCMS). In this system the particle
pairs are emitted transverse ($x=out$) to the reaction axis ($z=long$).
In Fig.2 we present the distributions of the relative $x$-coordinates
of the emission points and of the relative emission times calculated
in RQMD. The distributions of the relative $y$- and $z$-coordinates
show practically no asymmetry as expected from symmetry reasons
(the azimuthal symmetry and the symmetry of the initial system together
with the choice of a symmetric mid--rapidity window).

As pointed out in \cite{led96}, we can access the information about these
asymmetries exploiting the fact that the final state interaction,
determining the correlation function of two non--identical particles,
depends on the orientation of the relative 4--coordinates of the emission
points through the scalar product $\vec{q}\cdot\vec{r}^*$, where
$\vec{r}^*=\vec{r}_1^*-\vec{r}_2^*\equiv \{\Delta x^*,\Delta y^*,\Delta z^*\}$
and $\vec{q}=\vec{p}_1^*=-\vec{p}_2^*$ are the relative 3--coordinates
of the emission points and half the relative momentum in the two--particle
c.m.s. Since this system moves in the x-direction with the velocity
$v_{\perp}=|\vec{p}_{1\perp}+\vec{p}_{1\perp}|/(E_{1}+E_{2})$
and the Lorentz factor $\gamma_{\perp}=(1-v_{\perp}^2)^{-1/2}$, we have:
\begin{equation}
\Delta x^*=\gamma_{\perp}(\Delta x-v_{\perp}\Delta t),~~\Delta y^*=\Delta y,~~
\Delta z^*=\Delta z.
\end{equation}
Since further
$\langle \vec{q}\cdot\vec{r}^*\rangle =q_x\langle \Delta{x}^*\rangle
+q_y\langle \Delta{y}^*\rangle +q_z\langle \Delta{z}^*\rangle
$
and, in our case $\Delta{y}=\Delta{z}=0$,
we can search for the LCMS space (out) and time asymmetries
with the help of the correlation functions ${\cal R}_+$ and
${\cal R}_-$ corresponding to $q_x=\vec{q}\cdot\vec{v}_{\perp}/v_{\perp} > 0$ and
$< 0$, respectively \cite{led96}. In the case when the time differences
dominate over
the spatial ones, i.e. when $v|t|\gg r$ on average, the ratio of the
correlation functions
${\cal R}_{+}$ and ${\cal R}_{-}$ directly measures the mean relative emission time,
including its sign \cite{led96}.
Noting that for particles of equal masses the sign of the scalar product
$\vec{q}\cdot\vec{v}$ coincides with the sign of the velocity difference
$v_1-v_2$, we can see the simple classical meaning of the above selection.
It corresponds to the intuitive expectation of different particle interaction
in the case when the faster particle is emitted earlier as compared to the case
of its later emission. In former case the interaction between the two particles
will be weaker and the correlation function ${\cal R}_+$ will be closer to unity than
${\cal R}_-$.

To clarify the origin of the asymmetry effect, we will follow the arguments
given in \cite{led96}.
On the usual assumptions of a small particle phase-space density
and sufficiently smooth behavior of the single-particle spectra,
the two--particle
correlation function is determined by the modulus squared of the two-particle
amplitude $\psi_{-\vec{q}}(\vec{r}^*)$ averaged over the relative coordinates
of the emission points in the two-particle c.m.s. \cite{led82}:
\begin{equation}
R(p_1,p_2)=\langle|\psi_{-\vec{q}}(\vec{r}^*)|^2\rangle .
\end{equation}
For charged particles the final state interaction is often dominated by the
Coulomb interaction so that
\begin{equation}
\psi_{-\vec{q}}(\vec{r}^*)={\rm e}^{i\delta_0^c}
\sqrt{A_c(\eta )}{\rm e}^{-i\vec{q}\vec{r}^*}F(-i\eta ,1,i\xi ),
\end{equation}
where $\xi =qr^{*}+\vec{q}\vec{r}^*\equiv  qr^{*}(1+\cos {\theta ^{*}})$,
$\eta =1/(k^{*}a)$,
$a=1/(\mu z_1z_2e^2)$ is the Bohr
radius ($\mu $ is the reduced mass of the two particles, $z_1z_2e^2$ is the
product of the particle electric charges;
$a=$ 248.5, 222.5 and 83.6 fm for the $\pi^+p$, $K^+p$ and $\pi+K^+$
systems, respectively),
$\delta _0^c={\rm arg}\Gamma (1+i\eta )$ is the Coulomb s--wave phase shift,
\begin{equation}
\label{cfac}A_c(\eta )=2\pi \eta [\exp (2\pi \eta )-1]^{-1}
\end{equation}
is the Coulomb factor - modulus squared of the non-relativistic Coulomb wave
function at zero distance - introduced already by Fermi in his theory of
$\beta $--decay,
\begin{equation}
\label{Fhyper}F(\alpha ,1,z)=1+\alpha z+\alpha (\alpha +1)z^2/2!^2+\alpha
(\alpha +1)(\alpha +2)z^3/3!^2+\cdots
\end{equation}
is the confluent hypergeometric function.
At small relative momenta $qr^*<1$ and $r^*\ll |a|$, we can then write
\begin{eqnarray}
\label{cf}
R(p_1,p_2) &=&A_c(\eta )\left [ 1+2\langle r^{*}
(1+\cos {\theta ^{*}})\rangle /a
+\cdots \right ]
\nonumber \\
 & \equiv & A_c(\eta )\left [ 1+2\langle r^{*}\rangle /a +
2\cos {\psi }\cdot \langle \Delta x^*\rangle /a
+2\sin {\psi }\cdot (\sin {\phi}\cdot \langle \Delta y^*\rangle
+\cos {\phi}\cdot \langle \Delta z^*\rangle )/a
+\cdots \right ] ,
\end{eqnarray}
where $\psi $ and $\phi $ are the polar and azimuthal angles of the vector
$\vec{q}$ with respect to the velocity vector ($x$-axis):
$\vec{q}=q\{\cos{\psi},\sin{\psi}\sin{\phi},\sin{\psi}\cos{\phi}\}$
and $\cos {\psi }=\vec{q}\vec{v}_{\perp}/(qv_{\perp})$.
Noting that for particles with the same (opposite) charges the Bohr
"radius" $a > 0$ ($< 0$) and the Coulomb factor
$A_c(\eta )$ approaches unity from below (above),
we can immediately see that the interaction
(and thus the correlation) is stronger if
$\langle \vec{q}\vec{r}^*\rangle = \langle qr^*\cos {\theta ^{*}}\rangle <0$.
In the classical limit of $qr^*\gg 1$
this corresponds to the expectation of stronger interaction between particles
moving after the emission
(on average) towards each other in their c.m.s.\footnote
{
It should be noted that the two--particle amplitude
$\psi_{-\vec{q}}(\vec{r}^*)$ is often misidentified with the usual
two-particle
wave function in the scattering problem
$\psi_{\vec{q}}(\vec{r}^*)$ (see, e.g., \cite{koo77}).
This misidentification would obviously lead to the opposite asymmetry
effect in the correlation function.
}

Averaging in Eq. (\ref{cf}) over nearly isotropic distribution
of the vector $\vec{q}$ at $q\rightarrow 0$
(nearly uniform $\cos\psi$- and $\phi$--distributions)
or recalling that in the considered case $\langle \Delta{y}^*\rangle =
\langle \Delta{z}^*\rangle = 0$,
we get for the ratio
${\cal R}_+/{\cal R}_-$ at $q\rightarrow 0$
\begin{equation}
\label{ratapp}
{\cal R}_+/{\cal R}_-\approx 1+2\left(
\langle \cos {\psi }\rangle _+ -\langle \cos {\psi }\rangle _-\right)
\langle \Delta x^*\rangle /a,
\end{equation}
where the $\pm $ subscripts correspond to the averaging over
positive/negative values of $q_x=q \cos {\psi }$. The result of this averaging
depends on the selected kinematic region. For example, if only pairs
with the relative momentum $\vec{q}$ parallel or antiparallel to the
pair velocity are selected, then $\langle \cos {\psi }\rangle _{\pm}=\pm 1$
and the measured asymmetry would be maximal \cite{led96}.
In the case of averaging over a full range of
the uniform $\cos {\psi }$-distribution,
we have  $\langle \cos {\psi }\rangle _{\pm}=\pm 1/2$ and
twice less the asymmetry.
For our kinematical selection, the corresponding angular factor in
brackets in Eq. (\ref{ratapp}) is somewhat higher than unity.

Regarding the effect of the strong final state interaction,
since the magnitude of the two-particle scattering amplitude $f$
is usually much less than the mean $r^*$, it only slightly modifies
the correlation functions ${\cal R}_+$, ${\cal R}_-$ and their ratio at small $q$.
In particular, Eq. (\ref{ratapp}) for the ratio ${\cal R}_+/{\cal R}_-$ at
$q \rightarrow 0$ is modified by the substitution
$\langle \Delta x^*\rangle \rightarrow \langle \Delta x^*\rangle +{\rm Re}f
\cdot\langle \Delta x^*/r^*\rangle $. This means that for neutral particles
(when $|a| \rightarrow \infty $) the ratio ${\cal R}_+/{\cal R}_- \rightarrow 1$
at $q \rightarrow 0$.\footnote
{In \cite{led96} it was not clearly stated that the asymmetry
vanishes at $q \rightarrow 0$  only on the absence of the Coulomb
interaction.
}

The correlation functions ${\cal R}_+$, ${\cal R}_-$ and their ratio are plotted in Fig. 3.
We can see that for $\pi ^+p$ and $\pi ^+K^+$ systems these ratios are
less than unity
at small values of $q$, while for $K^+p$ system the ratio ${\cal R}_+/{\cal R}_-$
practically coincides with unity. These results well agree with the mean
values
of $\Delta t$,  $\Delta x$ and $\Delta x^*$ presented in Table 1.
We may see from Eqs. (1) and (\ref{ratapp})
that the absence of the effect in the ${\cal R}_+/{\cal R}_-$ ratio for the  $K^+p$
system is due to practically the complete compensation of the space and time
asymmetries leading to $\Delta x^* \approx 0$. For $\pi^+p$ system the effect
is determined mainly by the x-asymmetry. For $\pi^+K^+$ system both the
x- and time-asymmetries contribute in the same direction,
the latter contribution
being somewhat larger.

We have thus demonstrated that the unlike particle correlations allow
to directly
access the information about space-time asymmetries of particle production
in ultrarelativistic heavy ion collisions
even at present energies. The separation of the relative time delays from the
spatial asymmetry is, in principle, possible (see Eq. (1))
by studying the ratio ${\cal R}_+/{\cal R}_-$ in different intervals of the pair velocity.
However, such a study can hardly be done in a model independent way
due to the velocity dependence of the out asymmetry $\Delta x$
(particularly, both $\Delta x$ and $v_{\perp}\Delta t$ vanish at zero
LCMS pair velocity $v_{\perp}$).


\begin{table}[htbp]
\caption[]{\footnotesize
Mean values of the relative space-time coordinates in LCMS
(in fm) calculated from
RQMD (v2.3) for 158 A$\cdot$GeV $Pb+Pb$ central collisions.
}
\label{trq}
\medskip
\begin{center}
\begin{tabular}{|c|c|c|c|c|c|}
\hline
     &system & $\langle\Delta t\rangle$ & $\langle\Delta x\rangle$
     & $\langle\Delta x -v_{\perp}\Delta t\rangle$
     & $\langle\Delta x^*\rangle$\\
\hline
        &$\pi^+p$   &-0.5 &-6.2 &-6.4 &-7.9\\
        &$\pi^+K^+$ & 4.8 &-2.7 &-5.8 &-7.9\\
        &$K^+p$    &-5.5 &-3.2 &-0.6 &-0.4\\
\hline
\end{tabular}
\end{center}
\end{table}

\begin{figure}
\epsfig{file=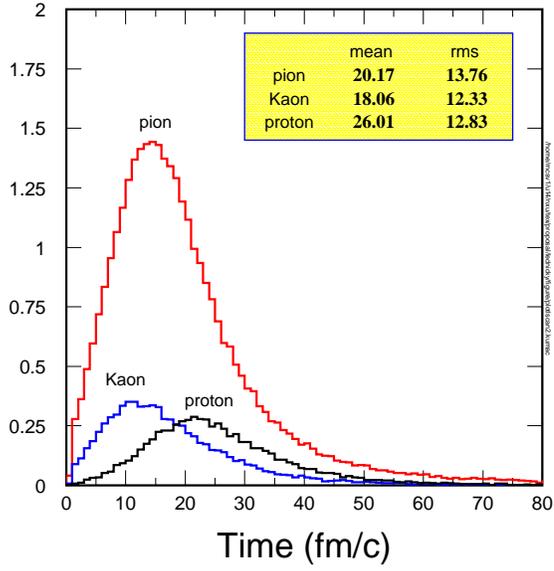,width=10.0cm}
\caption{Emission profiles for mid-rapidity
($|y|<1$) positive pions, kaons
and protons from 158 A$\cdot$GeV/c Pb+Pb central collisions. }
\label{fig1}
\end{figure}

\begin{figure}
\epsfig{file=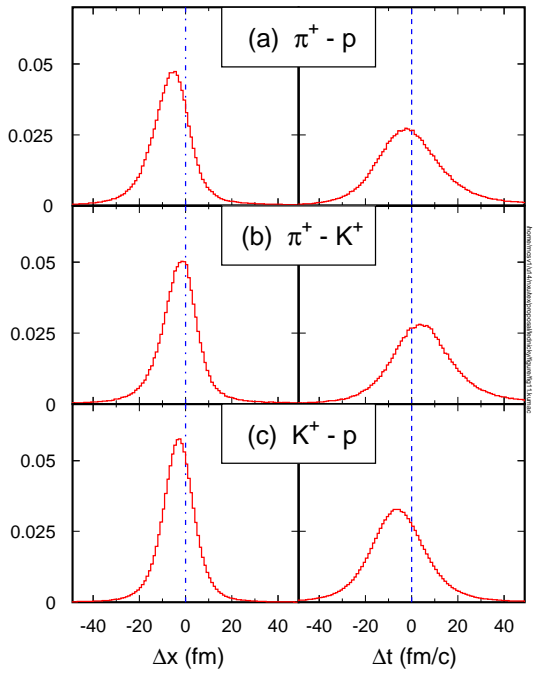,width=9.5cm}
\caption{Space-time distributions for
mid-rapidity particle
pairs $\pi^+ - p$, $\pi^+ - K^+$, and $K^+ - p$.}
\label{fig2}
\end{figure}

\begin{figure}
\epsfig{file=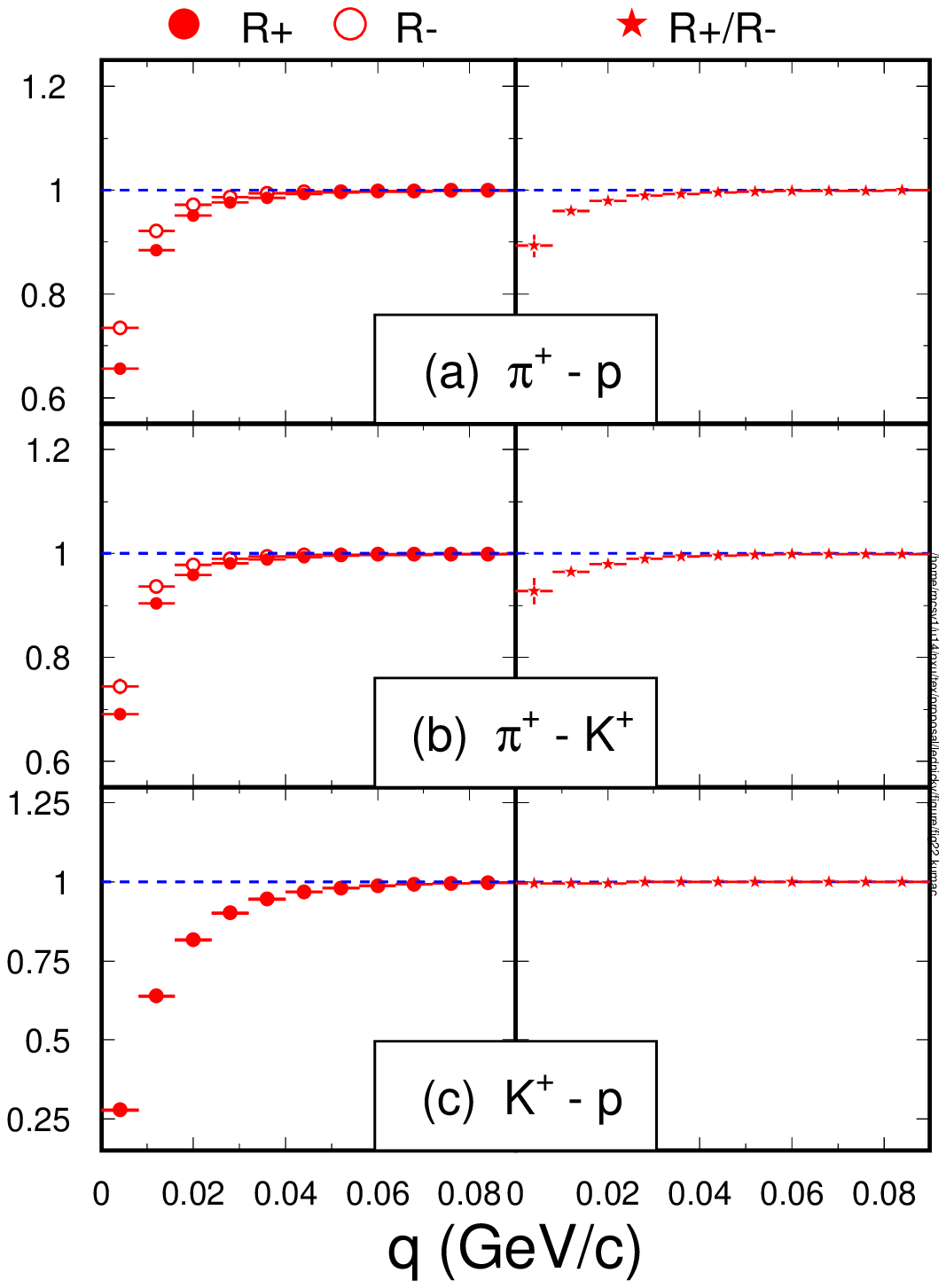,width=9.5cm}
\caption{Unlike particle correlation functions
for mid-rapidity particle
pairs $\pi^+ - p$, $\pi^+ - K^+$, and $K^+ - p$.}
\label{fig3}
\end{figure}



\begin{thebibliography}{999}

\bibitem{nag82} S. Nagamiya {\it et al.}, Phys. Rev. Lett. {\bf 49} (1982) 1383.

\bibitem{led82} R. Lednicky, V. L. Lyuboshitz, Sov. J. Nucl Phys. {\bf 35}
(1982) 770; Proc. Int. Workshop on Particle
Correlations and Interferometry in Nuclear Collisions, CORINNE 90, Nantes,
France, 1990 (ed. D. Ardouin, World Scientific, 1990) p. 42.; Heavy Ion Phys.
{\bf 3} (1996) 1.

\bibitem{gmi86} M. Gmitro, J. Kvasil, R. Lednicky,
V.L. Lyuboshitz, Czech. J. Phys. {\bf B36} (1986) 1281.

\bibitem{led96}  R. Lednicky, V.L. Lyuboshitz, B. Erazmus, D. Nouais,
Phys. Letters {\bf B373} (1996) 30.;
Report-94-22, Nantes, 1994.

\bibitem{koo77}  S.E. Koonin, Phys. Letters {\bf B70} (1977) 43; D.J. Dean,
S.E. Koonin, Phys. Letters {\bf B305} (1993) 5.

\end{thebibliography}
\end{document}